\documentclass[twocolumn,english,aip,jcp,reprint]{revtex4-1}
\usepackage[T1]{fontenc}
\usepackage[latin9]{inputenc}
\usepackage{color}
\usepackage{amsmath}
\usepackage{amssymb}
\usepackage{graphicx}

\makeatletter
\usepackage{times}

\renewcommand\[{\begin{equation}}
\renewcommand\]{\end{equation}}

\makeatother

\usepackage{babel}
\begin{document}
\global\long\def\avg#1{\langle#1\rangle}

\global\long\def\p{\prime}

\global\long\def\ket#1{|#1\rangle}

\global\long\def\bra#1{\langle#1|}

\global\long\def\proj#1#2{|#1\rangle\langle#2|}

\global\long\def\inner#1#2{\langle#1|#2\rangle}

\global\long\def\tr{\mathrm{tr}}

\global\long\def\dg{\dagger}

\global\long\def\im{\imath}

\global\long\def\pd#1#2{\frac{\partial#1}{\partial#2}}

\global\long\def\spd#1#2{\frac{\partial^{2}#1}{\partial#2^{2}}}

\global\long\def\der#1#2{\frac{d#1}{d#2}}

\global\long\def\l{\mathcal{L}}

\global\long\def\r{\mathcal{R}}

\global\long\def\s{\mathcal{S}}

\global\long\def\u{\mathcal{U}}

\global\long\def\v{\mathcal{V}}

\global\long\def\Ie{I_{\mathrm{exact}}}

\global\long\def\Is{I_{\mathrm{sim}}}

\global\long\def\corr{a_{k}^{\dg}b_{k^{\p}}\hspace{-0.8mm}\left(t\right)}

\title{Gibbs phenomenon and the emergence of the steady-state in quantum
transport}

\author{Michael Zwolak}
\email{mpz@nist.gov}

\affiliation{Biophysics Group, Microsystems and Nanotechnology Division, Physical
Measurement Laboratory, National Institute of Standards and Technology,
Gaithersburg, MD 20899, USA}
\begin{abstract}
Simulations are increasingly employing explicit reservoirs -- internal,
finite regions -- to drive electronic or particle transport. This
naturally occurs in simulations of transport via ultracold atomic
gases. Whether the simulation is numerical or physical, these approaches
rely on the rapid development of the steady state. We demonstrate
that steady state formation is a manifestation of the Gibbs phenomenon
well-known in signal processing and in truncated discrete Fourier
expansions. Each particle separately develops into an \emph{individual}
steady state due to the spreading of its wave packet in energy. The
rise to the steady state for an individual particle depends on the
particle energy -- and thus can be slow -- and ringing oscillations
appear due to filtering of the response through the electronic bandwidth.
However, the rise to the total steady state -- the one from all particles
-- is rapid, with timescale $\pi/W$, where $W$ is the bandwidth.
Ringing oscillations are now also filtered through the bias window,
and they decay with a higher power. The Gibbs constant -- the overshoot
of the first ring -- can appear in the simulation error. These results
shed light on the formation of the steady state and support the practical
use of explicit reservoirs to simulate transport at the nanoscale
or using ultracold atomic lattices.
\end{abstract}
\maketitle

An increasing number of nanoscale electronic \citep{Tao06-1,Song11-1,Lortscher13-1,ratner_brief_2013,sun_single-molecule_2014,wang_modulation_2017}
studies aim at probing and exploiting dynamical phenomena at both
slow and fast timescales \citep{Platero04-1,Kohler05-1,Feve07-1,Zhong08-1,Ward08-1,Terada10-1,Marquardt11-1,moore_polarizabilities_2010,mullegger_radio_2014,trasobares_17_2016,cocker_tracking_2016}.
Moreover, finite, closed ultra-cold atomic systems\citep{giorgini_theory_2008,bloch_many-body_2008}
simulate transient transport \citep{Schneider12-1,Chien12-1,Brantut12-1,Chien13-1,Chien14-1,krinner_observation_2015,chien_quantum_2015,krinner_two-terminal_2017,gruss_energy-resolved_2018}
and can examine the generation of topological matter via time-dependent
fields \citep{eckardt_colloquium_2017,weinberg_adiabatic_2017}. An
avenue to computationally study transient and dynamical phenomena
is to include particle reservoirs explicitly in the simulation, essentially
letting a ``capacitor'' discharge and drive current through a region
of interest \citep{Zwolak04-1,DiVentra05-1,Bushong05-1,Schneider06-1,Schmitteckert06-1,Cheng06-1,Hassanieh06-1,Bushong07-1,Sai07-1,Dias08-1,Evans09-1,Eshuis09-1,Heidrich09-1,myohanen_kadanoff-baym_2009,kurth_dynamical_2010,Branschadel10-1,Wang11-1,Varga11-1,gaury_numerical_2014}.
The inclusion of relaxation can give a true steady state while still
permitting the examination of transient/dynamical processes \citep{gruss_landauers_2016,gruss_communication:_2017,elenewski_communication:_2017}
(including for thermal transport \citep{velizhanin_crossover_2015,chien_thermal_2017,chien_topological_2018}).
This type of ``open'' system approach has a long history \citep{sanchez_molecular_2006,subotnik_nonequilibrium_2009}
(see discussion in Ref.~\onlinecite{elenewski_communication:_2017}),
including designs for time-dependent density functional theory (TD-DFT)
\citep{zelovich_state_2014,zelovich_parameter-free_2017,morzan_electron_2017}.
However, large-scale numerical simulations (e.g., integrating correlation
matrices, numerical renormalization and tensor network methods, TD-DFT,
or other techniques) generally do not give direct insight into the
formation of the steady state and the factors controlling transient
behavior.

\begin{figure}
\begin{centering}
\includegraphics[width=1\columnwidth]{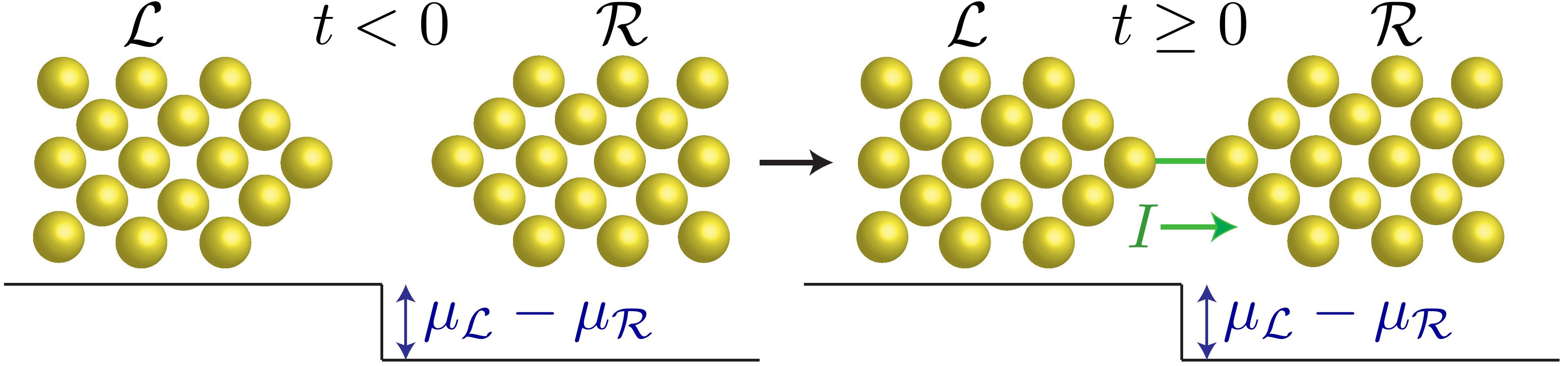}
\par\end{centering}
\caption{Schematic of a lattice set out of equilibrium by adding a link at
time $t=0$ between two initially disconnected regions $\protect\l$
and $\protect\r$. There is a density imbalance (black line) that
creates a chemical potential drop $\mu_{\protect\l}-\mu_{\protect\r}$
(alternatively, there can initially be a uniform potential and a bias
simultaneously turns on when connecting the lattice). The current,
$I$, is the step response to the addition of the link (green) filtered
by the electronic bandwidth and bias window.\label{fig:Schematic}}
\end{figure}

Here, we employ a Kubo approach to study transients in closed, noninteracting
fermionic systems. We demonstrate its application using a system
set out of equilibrium by connecting initially disjoint lattices,
see Fig. \ref{fig:Schematic}, a technique related to the tunneling
Hamiltonian and Green's function approaches to transport. We show
how the steady state arises, how oscillations decay, and how different
frequency scales contribute to transpor\textcolor{black}{t, as well
quantify aspects of simulation error. }We expect that this approach
will find application in dynamical, many-body transport in both nanoscale
and ultracold atomic systems, including diagnosing pathological numerical
setups and increasing simulation efficiency. 

Before the lattices come into contact (i.e., for times $t<0$ in Fig.
\ref{fig:Schematic}), the Hamiltonian is 
\[
H_{0}=H_{\l}+H_{\r},
\]
with 
\[
H_{\l}=\sum_{k\in\l}\hbar\omega_{k}a_{k}^{\dg}a_{k},\,H_{\r}=\sum_{k\in\r}\hbar\omega_{k}b_{k}^{\dg}b_{k}
\]
and $a_{k}^{\dg}$ ($a_{k}$) and $b_{k}^{\dg}$ ($b_{k}$) are the
fermionic creation (annihilation) operators on the left ($\l$) and
right ($\r$), respectively. These are noninteracting lattices with
$N_{\l\left(\r\right)}$ levels and frequencies $\omega_{k}$. The
initial state is one with a density imbalance, where the left region
has particles up to the chemical potential $\mu_{\l}$ and the right
to $\mu_{\r}$. This drives the current when, at $t=0$, the perturbing
Hamiltonian 
\[
H^{\p}=\sum_{k\in\l,k^{\p}\in\r}\hbar v_{kk^{\p}}\left(a_{k}^{\dg}b_{k^{\p}}+b_{k^{\p}}^{\dg}a_{k}\right)
\]
connects the two lattices, as shown in Fig. \ref{fig:Schematic}.
The strength of the connection is the total hopping frequency $v=\sqrt{\sum_{k,k^{\p}}v_{kk^{\p}}^{2}}$,
which we will treat as a perturbation.\textcolor{black}{{} The density
imbalance encodes the chemical potential in the initial state, making
the calculations non-perturbative in the bias}\textcolor{red}{}\footnote{\textcolor{black}{For instance, the approach captures negative differential
conductance when a bias is switched on simultaneously with the weak
link. There is, of course, a change in the chemical potential with
time due to having finite systems, which is not captured by the approach.}} (\textcolor{black}{unlike, e.g., Refs.~\onlinecite{bohr_dmrg_2006,bohr_strong_2007,schmitteckert_exact_2008},
which employ numerical renormalization in tandem with a Kubo approach
with the applied bias as the perturbation}). We can relate this to
a real-space model with contact at, e.g., one site via the identification
$c_{1}=\sum_{k}\u_{1k}a_{k}$, $d_{1}=\sum_{k}\v_{1k}b_{k}$ and $v_{kk^{\p}}=v\,\u_{k1}^{\star}\v_{1k^{\p}}$,
giving the connection $\hbar v\left(c_{1}^{\dg}d_{1}+d_{1}^{\dg}c_{1}\right)$.
Here, the quantities $\u$ and $\v$ are the transformation matrices
from energy- to real-space on the left and right lattices. 

We will apply the Kubo formula
\begin{equation}
\avg{O\left(t\right)}=\avg O_{0}-\im\int_{0}^{t}dt^{\p}\avg{\left[O\left(t\right),H^{\p}\left(t^{\p}\right)\right]}_{0}\label{eq:Kubo}
\end{equation}
for the observable $O$, where $O(t)=e^{\im H_{0}t}Oe^{-\im H_{0}t}$
is an operator in the interaction picture and $\avg O_{0}$ indicates
an average with respect to the initial state. While our focus is on
closed, finite systems, we will take the infinite system limit to
make some expressions more transparent. This will not obscure their
interpretation for finite systems. 

The particle current from left to right is
\[
I(t)=-\avg{dN_{\l}/dt}=-2\sum_{k,k^{\p}}v_{kk^{\p}}\Im\avg{\corr}
\]
for $t\ge0$. Here, $N_{\l}$ is the number operator in the Heisenberg
picture on the left, $dN_{\l}/dt=-\imath\left[N_{\l},H_{0}+H^{\p}\right]$,
and the factor of 2 appears due to taking the imaginary component
$\Im$ (i.e., not due to spin). Applying Eq. \eqref{eq:Kubo} to $\corr$
yields
\begin{equation}
\avg{\corr}=-v_{kk^{\p}}\left(n_{k}-n_{k^{\p}}\right)\frac{e^{\im t\left(\omega_{k}-\omega_{k^{\p}}\right)}-1}{\omega_{k}-\omega_{k^{\p}}},\label{eq:IkkpKubo}
\end{equation}
where $n_{k}$ are the initial particle occupancies and we use that
$\avg{a_{k}^{\dg}b_{k^{\p}}}_{0}=0$ for two initially disjoint lattices.
The total current from this perturbative result is thus
\begin{equation}
I(t)=2\sum_{k,k^{\p}}v_{kk^{\p}}^{2}\left(n_{k}-n_{k^{\p}}\right)\frac{\sin\left[\left(\omega_{k}-\omega_{k^{\p}}\right)t\right]}{\omega_{k}-\omega_{k^{\p}}}.\label{eq:CurrentFinite}
\end{equation}
So far we only assume that the two lattices are initially disconnected
and have occupancies from their separate single-particle eigenstates.
\begin{figure}
\begin{centering}
\includegraphics[width=1\columnwidth]{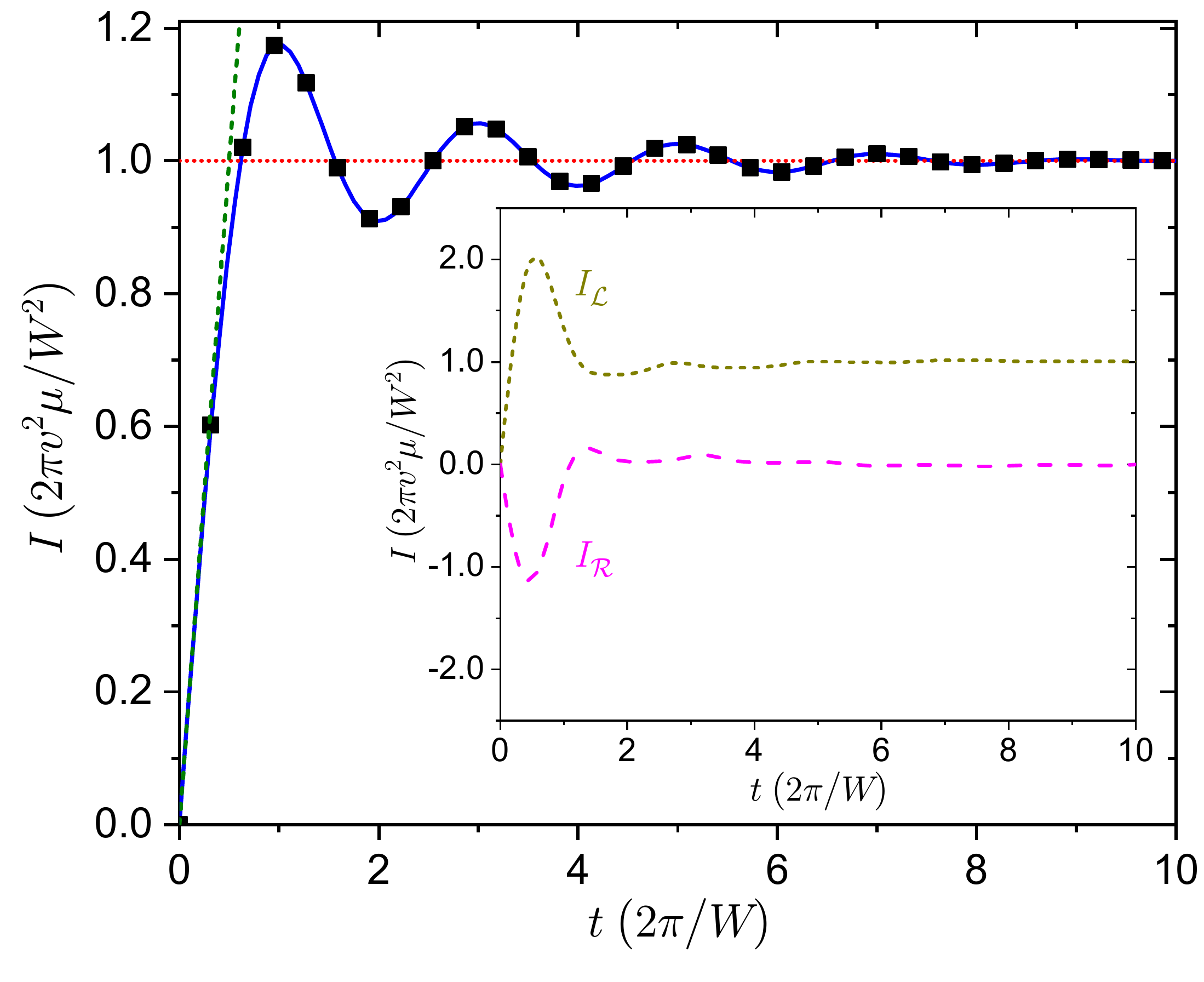}
\par\end{centering}
\caption{Current through a weak link when density-imbalanced flat band lattices
come into contact. The blue line (black squares) shows the Kubo (exact)
result for $\upsilon=W\cdot10^{-3}$, $\mu=W/10$, and $N_{\protect\l\left(\protect\r\right)}=100$.
The dotted red line is the steady state current and the dashed green
line the rise to the steady state at time $\pi/W$. Since this is
a finite system, the resultant current goes into a ``quasi''-steady
state. The inset shows the forward, $I_{\protect\l}$ {[}$n_{k}=1$
and $n_{k^{\protect\p}}=0$ in Eq. \eqref{eq:CurrentFinite}{]}, and
backward, $I_{\protect\r}$ {[}$n_{k}=0$ and $n_{k^{\protect\p}}=1$
in Eq. \eqref{eq:CurrentFinite}{]} currents. These have significantly
larger transients but they partially cancel, leaving more regular
-- but algebraically decaying -- oscillations in $I$. A true steady
state will form when $N\to\infty$ and then $t\to\infty$. \label{fig:Current} }
\end{figure}

Let's first examine the current, $I_{k}(t)$, from a particle in state
$k$ on the left going into an empty reservoir of bandwidth $W$ on
the right. Setting $v_{kk^{\p}}=v/\sqrt{N_{\l}N_{\r}}$ -- i.e.,
a flat band -- and taking $\sum_{k^{\p}}1/N_{\r}\to\int d\omega/W$,
gives 
\begin{equation}
I_{k}\left(t\right)=\frac{2v^{2}}{N_{\l}W}\int_{-W/2}^{W/2}d\omega^{\p}\,\frac{\sin\left[\left(\omega_{k}-\omega^{\p}\right)t\right]}{\omega_{k}-\omega^{\p}}.\label{eq:Ik}
\end{equation}
When $t\to\infty$, the integrand approaches $\pi\delta\left(\omega_{k}-\omega^{\p}\right)$,
expressing conservation of energy in the absence of inelastic processes
and in the long-time limit. This indicates the presence of a steady
state current of $2v^{2}\pi/N_{\l}W$ even if $N_{\l}=1$ (and, since
there is only one particle, it can be a fermion or a massive boson).
However, the perturbative expression does not capture that there is
a decay time\footnote{If the coupling from state $k$ to $\r$ was given by the spectral
function $\gamma/2\pi$ (i.e., flat but unbounded when $W\to\infty$),
then $n_{k}\left(t\right)$ will be exponentially decaying even for
short times \citep{Zwolak08-2}.} $T^{\star}=N_{\l}W/2v^{2}\pi$. For times much shorter than this,
the particle looks to be in a steady state. This demonstrates that
constructive contributions from many incoming particles are not necessary
for steady state formation, but rather it is the spread of a single
particle into many different states -- its wave-like nature in energy
space -- that results in a nearly steady current. Since only a single
particle is present, the steady state is just a linear increase in
time of the probability for the particle to be in $\r$, which is
possible to measure in cold atom lattices by repetition of the experiment
many times. 

For finite times, the integral in Eq. \eqref{eq:Ik} is just 
\begin{equation}
I_{k}\left(t\right)=\frac{2v^{2}}{N_{\l}W}\left\{ \mathrm{Si}\left[t(\omega_{k}+W/2)\right]-\mathrm{Si}\left[t(\omega_{k}-W/2)\right]\right\} ,\label{eq:Ik2}
\end{equation}
where $\mathrm{Si}\left[\circ\right]$ is the sine integral. The derivative
$\left.dI/dt\right|_{t=0}$ determines the rise to the steady state.
For the single particle, this depends on the smaller of the two energies,
$\left|\omega_{k}+W/2\right|$ or $\left|\omega_{k}-W/2\right|$.
For instance, for $\omega_{k}=0$, the initial (linear) rise occurs
with slope $2v^{2}/N_{\l}$. Thus, the time to reach the steady state
value, $2v^{2}\pi/N_{\l}W$, is $\pi/W$, at which time the current
begins oscillating. If $\omega_{k}$ (in $\l$) approaches the band
edge (in $\r$), then the steady state takes a long time to develop.
In that case, there is a fast process -- where one of the sine integral
quickly rises -- and a slow process -- where the other rises with
time $\sim1/\left(W/2-\left|\omega_{k}\right|\right)$. After the
initial rise, oscillations -- ringing -- appear, which decay as
the steady state is approached which decay as the steady state is
approached. Such oscillations are seen in extended reservoir, microcanonical,
and related approaches, in addition to numerical integration of the
time-dependent Green\textquoteright s functions \citep{Jauho94-1}. 

For $\omega_{k}=0$, the rapid rise and ``ringing oscillation''
is none other than Gibbs phenomenon \citep{wilbraham_certain_1848,gibbs_fouriers_1898,gibbs_fouriers_1899,bocher_introduction_1906,hewitt_gibbs-wilbraham_1979}
for the step function $\mathrm{sign}\left[t\right]$ sent through
a low-pass frequency filter. The Fourier transform of $\mathrm{sign}\left[t\right]$
is $\im\sqrt{2/\pi}/\omega$. Filtering the frequencies outside of
the bandwidth $\left[-W/2,W/2\right]$ and taking the inverse transform
gives Eq. \eqref{eq:Ik} up to a prefactor \footnote{The contribution from the cosine component of the inverse Fourier
transform is zero. However, working directly with Eq. \eqref{eq:IkkpKubo},
i.e., without first taking the imaginary component, the integral of
$(\cos\left[t\left(\omega_{k}-\omega^{\p}\right)\right]-1)/(\omega_{k}-\omega^{\p})$
yields a non-zero real component. This is not part of the signal of
interest. We note that the $v^{2}$ contribution to the prefactor
can be written as $v^{2}\Theta\left(t\right)$, which properly removes
the signal for $t<0$. }. The oscillations are thus an inherent aspect of electronic transport.
Moreover, the ``overshoot'' of the current -- its first and maximum
oscillation overtop the steady state value -- is $G\cdot2v^{2}\pi/NW$,
where $G=2\mathrm{Si}\left[\pi\right]/\pi-1=0.1789\ldots$ is the
Gibbs constant. That is, the overshoot is about 18 \% higher than
the steady state value. Regardless of the bandwidth, the magnitude
of the overshoot -- and, indeed, the dimensionless form of the current
-- stays the same. When examining $\omega_{k}\neq0$, these basic
insights remain but now the filter acts asymmetrically, introducing
oscillations that depend on both $W$ and $\omega_{k}$. Different
spectral densities of the reservoirs and strong coupling will give
different overshoot values. However, the physical process is universal,
the signal is filtered through the bandwidth giving rise to ringing
oscillations. 

We now examine the total current in the presence of a chemical potential
drop. Considering the flat band case and equal bandwidths in $\l$
and $\r$, the continuum limit of Eq. \eqref{eq:CurrentFinite} gives
\begin{equation}
I\left(t\right)=\int_{-W/2}^{W/2}d\omega\,\delta I\left(\omega,t\right),\label{eq:KuboFlatIntegral}
\end{equation}
where the contribution to the current at frequency $\omega$ in $\l$
is
\begin{equation}
\delta I=\frac{2v^{2}}{W^{2}}\int_{-W/2}^{W/2}d\omega^{\p}\,\left[n_{\l}\left(\omega\right)-n_{\r}\left(\omega^{\p}\right)\right]\frac{\sin\left[\left(\omega-\omega^{\p}\right)t\right]}{\omega-\omega^{\p}}.\label{eq:dIw}
\end{equation}
We now explicitly label the occupancies $n_{\l\left(\r\right)}$.
The steady state current is $2\pi v^{2}\mu/W^{2}$ for a chemical
potential drop of $\mu=\mu_{\l}-\mu_{\r}$. Equations \eqref{eq:KuboFlatIntegral}
and \eqref{eq:dIw} show that, to highest order in $v$, there is
a one way flow from filled states on the left into empty states on
the right lattice when $\mu_{\l}>\mu_{\r}$. Indeed, as with Eq. \eqref{eq:Ik},
states at frequency $\omega$ go into states $\omega^{\p}=\omega$
as $t\to\infty$, giving the standard bias window.
\begin{figure}
\begin{centering}
\includegraphics[width=1\columnwidth]{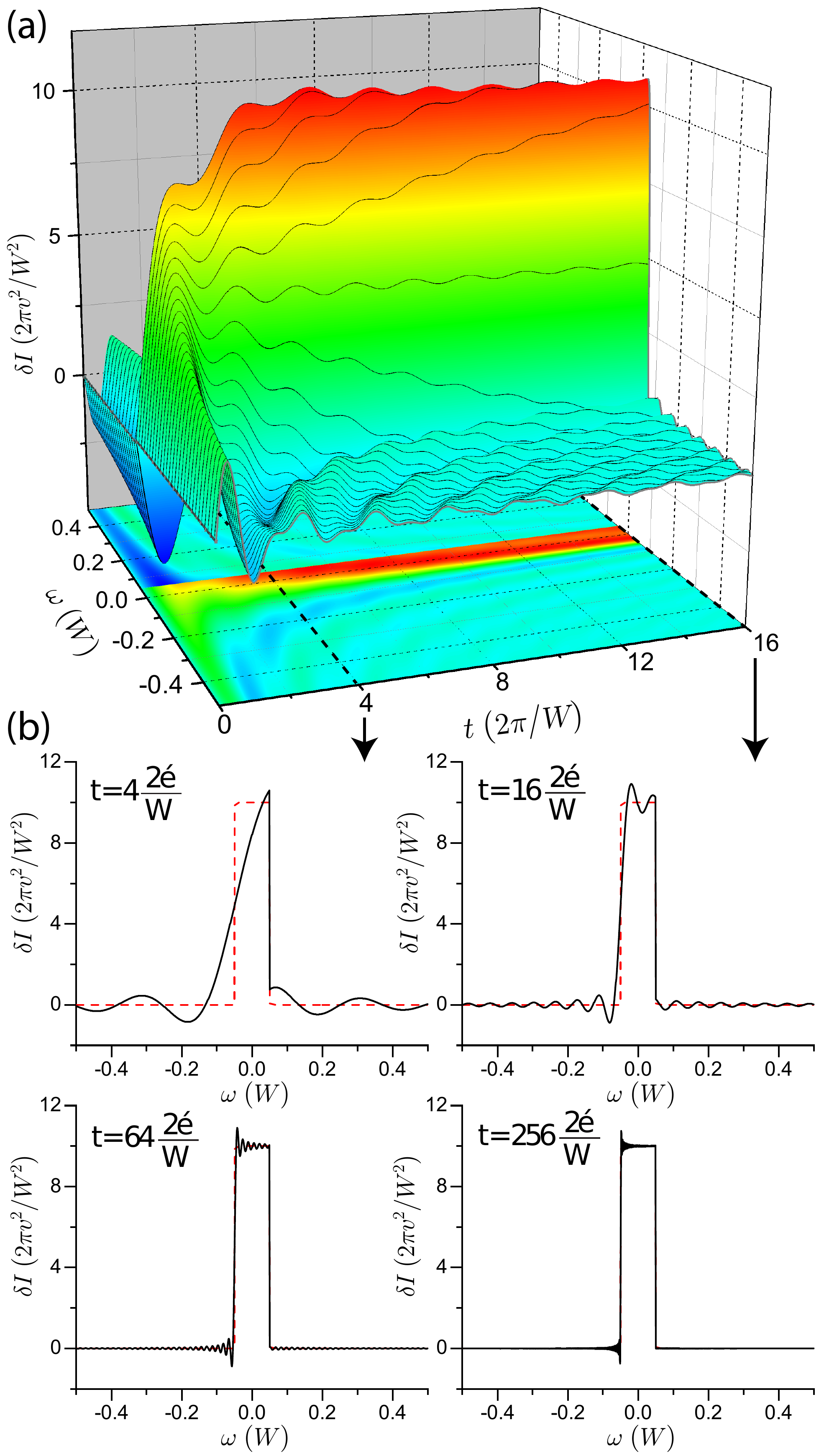}
\par\end{centering}
\caption{Current contribution, $\delta I$, from different frequencies in $\protect\l$.
(a) Initially, all states contribute substantially to the current,
but contributions above and below the Fermi level partially cancel,
see $I_{\protect\l}$ and $I_{\protect\r}$ in the Fig. \ref{fig:Current}
inset. (b) Contributions to the current for times $t$ indicated on
each panel. As $t\to\infty$, the contributions approach the red dashed
line, with vanishing values outside the bias window, $-\mu/2\le\omega\le\mu/2$.
The oscillatory features on the left side of the bias window reflect
the occurrence of Gibbs phenomenon. These oscillations do not disappear
but rather get squeezed toward the jump at the bias window edge. No
oscillations exist on the positive side due to the nature of that
edge (a cutoff from the Fermi-Dirac occupation). \label{fig:CurrentContrib}
}
\end{figure}

Taking $\mu_{\l}=\mu/2$, $\mu_{\r}=-\mu/2$, and performing the integrations
at zero temperature (so $n_{\l\left(\r\right)}=0$ or 1) yields 
\begin{align}
I\left(t\right) & =\frac{2v^{2}}{W^{2}}\left\{ \sum_{\pm}\pm(W\pm\mu)\mathrm{Si}\left[\frac{t}{2}(W\pm\mu)\right]\right.\nonumber \\
 & \left.+\frac{4\sin\left[\frac{Wt}{2}\right]\sin\left[\frac{\mu t}{2}\right]}{t}\right\} \label{eq:KuboFlat}\\
 & \approx\frac{4v^{2}}{W^{2}}\mathrm{Si}\left[Wt/2\right]\mu,
\end{align}
where the second expression is for a small bias, showing exactly the
same manifestation of Gibbs phenomenon as the individual particles
at the Fermi level. Figure \ref{fig:Current} shows the Kubo result,
Eq. \eqref{eq:KuboFlat}, together with the exact result for a finite
system, as well as the steady state value and initial rise. Just like
individual particles at the Fermi level, the total current rises with
time $\pi/W$. Unlike individual particles, this result is nearly
true even when a small frequency scale appears in Eq. \eqref{eq:KuboFlat},
e.g., $\left(W-\mu\right)$ for a chemical potential drop comparable
to the bandwidth. The component with the small frequency scale takes
a longer time to reach its steady state but it appears with a prefactor
that is also the small frequency scale. Hence, while it takes time
to rise, it has a small contribution to the total current. As a separate
note, the convergence to the infinite system limit is non-monotonic
purely due to the discrete nature of the states and fillin{\small{}g
}\footnote{The total coupling of modes in the bias window determines the current
magnitude. Each mode can be thought of taking up $\Delta=W/N$ width
of the spectral density. For even $N$ and placing the modes uniformly
across the total bandwidth ($-W/2+\Delta/2+(k-1)\Delta$, with $k=1,\ldots,N$),
the number of modes in the bias window is $n=2\cdot\mathrm{Floor}\left[N\mu/2W+1/2\right]$
when the bias is applied symmetrically ($\mu_{\l}=-\mu_{\r}=\mu/2$).
To obtain the proper steady state current, the fraction of the spectral
density in the bias window must be equal to the bias as a fraction
of the bandwidth, $n/N\equiv\mu/W$. After rearranging, $N\mu/2W=n/2$
or, in other words, $N\mu/2W$ has to be an integer ($n$ is even).
When $N\mu/2W$ is not an integer, the modes in the bias window have
$n/N\lessgtr\mu/W$ and therefore either they underestimate or overestimate
the current. Note, as well, that $N\mu/2W=\mathrm{Integer}$ is a
purely artificial constraint to get the right coupling strength (and
hence current) for finite-size systems. As $N\to\infty$, the corrections
away from integer values decay as $1/N$. While these results are
for the flat band model with equally spaced modes, the calculation
can be done in other scenarios as well (more complex band structures
and couplings, and potentially even out of the weak coupling limit
by checking finite-size convergence of the Greens functions). \textbf{}.}, which gives insight into behavior observed in density functional
theory calculations \citep{yang_variational_2002}.

We can also examine the contribution to the current from different
frequency scales on the left, Eq. \eqref{eq:dIw}. All frequency scales
contribute to the current for short times, see Fig. \ref{fig:CurrentContrib}a,
but this contribution decays with both frequency and time. By $t=4\cdot2\pi/W$,
the contribution is small outside the bias window and, as time progresses,
it takes on the form of the bias window, Fig. \ref{fig:CurrentContrib}b
(the contributions reflect the band structure/couplings and thus are
flat for the flat band model). When solving problems numerically,
one reduces continuum reservoirs/environments into a finite, discrete
number of components. The decay of the contribution with frequency
(outside the bias window) suggests routes to alternative coarse grainings
in frequency to enhance the simulation efficiency, as done in Ref.~\onlinecite{Zwolak08-2}.
The influence of different frequency scales will ultimately depend
on details of the model (e.g., the presence of interactions, etc.),
but we expect that the Kubo approach will help reveal the errors incurred
by various coarse grainings. We leave this for future studies and
instead focus on errors in estimating the steady-state value of the
current. 

The rise time of the current is rapid, indicating that already for
small system sizes and times one can get a reasonably accurate value
of the steady-state current (in the model here, taking the first maximum
as an estimate of the steady-state current would only give a relative
error of $G$, about 18 \%). The slowly (algebraically) decaying nature
of oscillations, though, influence the accuracy of further simulation.
From Eq. \eqref{eq:KuboFlat}, the asymptotic decay of the current
to its steady state is \footnote{Equation \eqref{eq:KuboFlat} has an explicit oscillatory term that
decays as $1/t$. However, this cancels the highest order (in $1/t$)
term from the sine integrals.} 
\begin{equation}
-\frac{1}{t^{2}}\frac{2v^{2}}{W^{2}}\frac{8W\cos\left[\frac{Wt}{2}\right]\sin\left[\frac{\mu t}{2}\right]-8\mu\cos\left[\frac{\mu t}{2}\right]\sin\left[\frac{Wt}{2}\right]}{W^{2}-\mu^{2}},\label{eq:CurrentDecay}
\end{equation}
compared with 
\[
-\frac{1}{t}\frac{2v^{2}}{N_{\l}W}\frac{W\cos\left[\omega_{k}t\right]\cos\left[\frac{Wt}{2}\right]+2\omega_{k}\sin\left[\omega_{k}t\right]\sin\left[\frac{Wt}{2}\right]}{(W/2)^{2}-\omega_{k}^{2}}
\]
from Eq. \eqref{eq:Ik2} for a single particle going into an empty
band. Both expressions are in the long-limit compared to all other
timescales (namely, $1/\mu$ and $1/\omega_{k}$, as well as $1/W$).
In the case of an infinitesimal bias ($1/\mu\to\infty$ before the
long time limit), one also gets oscillations that decay as $1/t$
(specifically, $-4u\cos(Wt/2)/Wt$, as with the single particle at
$\omega_{k}=0$).
\begin{figure}
\begin{centering}
\includegraphics[width=1\columnwidth]{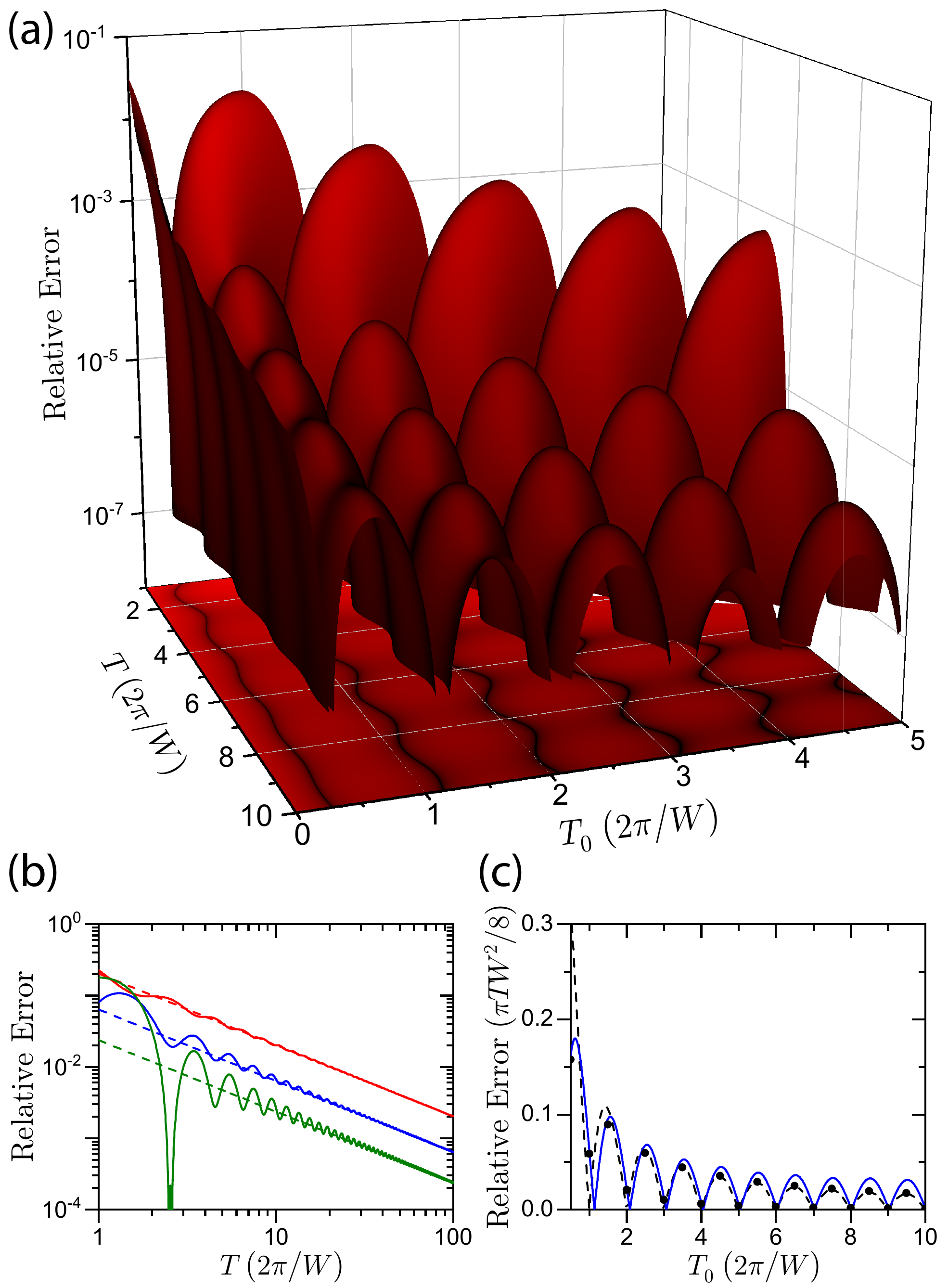}
\par\end{centering}
\caption{Relative error of the estimate, Eq. \eqref{eq:AvgCurrent}. (a) Error
versus $T$ and $T_{0}$. For a given total simulation time, $T$,
the error has minima when $T_{0}$ is approximately integer multiples
of $2\pi/W$. For long times (both $T$ and $T_{0}$), the oscillations
that can be seen on the 2D projection smooth to flat lines and the
minima approach these values. (b) Error versus $T$ for $T_{0}=0,$
$\pi/W,$ and $2\pi/W$ (solid red, blue, and green lines, respectively).
All of these decay as $1/T$ for large $T$, shown by the asymptotic
forms (dashed lines with the same colors). If $T_{0}$ is set to $3\pi/W$
(in between extrema), the error will be substantially larger than
when it is $2\pi/W$ (at an extremum). (c) Error normalized by $1/T$.
The blue line (black circles) shows $T=20\cdot2\pi/W$ ($200\cdot2\pi/W$).
The dashed, black line shows the asymptotic result, Eq. \eqref{eq:AsymTT0}.
So long as both $T$ and $T_{0}$ are large enough, the asymptotic
result captures the relative error. The minima in this limit are exactly
at integer multiples of $2\pi/W$ for $T_{0}$ regardless of $T$
(including non-integer multiples of $2\pi/W$). \label{fig:Errors}}
\end{figure}

\textcolor{black}{To obtain the steady state current, one has to deal
with finite size and finite time issues, including both real physical
effects (such as the decaying oscillations here) and artifacts (such
as persistent oscillations due to finite lattice sizes \citep{Branschadel10-1}).
One can remove persistent oscillations seen in impurity problems by
fitting \citep{Branschadel10-1}, reducing one source of uncertainty
but there are still other finite-size errors. In the case here, fitting
the decaying envelope of the oscillations to $1/T$ will give an accurate
estimate of the steady state current. However, complex models will
have much more complicated dynamics with oscillations at many time
scales and amplitudes, and potentially with a different decay in time.
Fitting, finding a bisecting line, or enveloping oscillations will
be difficult to implement when the oscillations and decay are more
irregular (although, a universal $1/T$ behavior appears in higher
cumulants \citep{schmitteckert_transport_2014,carr_full_2015}). We
thus assess ``model agnostic'' strategies -- strategies that do
not require specific knowledge of the model under study -- to obtain
the value of the steady state current that remove finite time effects,
the ones shown to be limiting in related contexts \citep{schmitteckert_transport_2014}. }

\textcolor{black}{Two agnostic strategies for e}stimating the steady
state current from a closed, finite-sized simulation are to (1) take
the value of the current at the end of the simulation or (2) average
the current over some region of time. These approaches sometimes serendipitously
yield the exact current. Thus, we will either work with error envelopes,
i.e., the smooth curve going through the set of maxima in the error
versus time, or with asymptotic forms for the error decay. Considering
the relative error, $1-\Is/\Ie$ with $\Is$ the current from simulation
and $\Ie$ the exact current, strategy (1) gives 
\[
\frac{8}{\pi\mu WT^{2}}
\]
for the \emph{error envelope}. Here, $T$ is the total simulation
time and we took $T\to\infty$ and then $\mu\to0$ (taking the limit
$\mu\to0$ and then $T\to\infty$ gives a leading $1/T$ decay in
the oscillations and error\footnote{The appropriate limit depends on the bias and time range of interest.
For $\mu=W/10$ and times past $10\cdot2\pi/W$, $T\to\infty$ should
be first.}). For strategy (2), the estimate is
\begin{equation}
\Is=\frac{1}{T-T_{0}}\int_{T_{0}}^{T}I\left(t\right)dt.\label{eq:AvgCurrent}
\end{equation}
Compact forms for the relative error follow from integrating this
equation with $I\left(t\right)$ from Eq. \eqref{eq:KuboFlatIntegral}.
To simplify calculations, we can work with the small bias expression
directly in the case of strategy (2), as the average in Eq. \eqref{eq:AvgCurrent}
will have a dominant error due to short time contributions. The error
will thus decay as $1/T$ so long as $T_{0}$ is not too large (i.e.,
either $T\to\infty$ and then $\mu\to0$, or the reverse, will do). 

Figure \ref{fig:Errors}a shows the relative error versus $T$ and
$T_{0}$ for strategy (2). The minimum error comes at approximately
integer multiples of $2\pi/W$ for $T_{0}$ -- at oscillatory extrema
of the current -- for \emph{any} value of $T$. Indeed, the asymptotic
error decay (first $T\to\infty$ and then $T_{0}\to\infty$),
\begin{equation}
\left|\frac{8\sin\left(T_{0}W/2\right)}{\pi W^{2}TT_{0}}\right|,\label{eq:AsymTT0}
\end{equation}
has minimal error exactly when $T_{0}$ is an integer multiple of
$2\pi/W$. The reason for this is that the integration in Eq. \eqref{eq:AvgCurrent}
accumulates excess error (e.g., $I\left(t\right)>I_{\mathrm{exact}}$)
before encountering terms (e.g., $I\left(t\right)<I_{\mathrm{exact}}$)
that cancel that excess. The maximal cancellation of errors occurs
when $T_{0}$ is at multiples of $2\pi/W$. If $T-T_{0}$ is a multiple
of $4\pi/W$ (i.e., a complete oscillation), then there are saddle
points on the error manifold when $T_{0}$ is at odd multiples of
$\pi/W$, but moving $T_{0}$ toward the extrema (holding $T$ constant)
decreases further the error. Figures \ref{fig:Errors}(b,c) show
the error decay for different $T_{0}$ and the coefficient of the
decay versus $T_{0}$. The asymptotic coefficient qualitatively captures
even the non-asymptotic regime. For small $T_{0}$, though, the coefficient
can be off in relative terms, which is not apparent on the scale of
Fig. \ref{fig:Errors}(c): Comparing Eq. \eqref{eq:AsymTT0} with
$T_{0}=2\pi/W$ to the actual decay, $2(2-\pi^{2}G)/\pi WT$, for
large $T$ but not large $T_{0}$, it is clear that the actual coefficient
of the decay is due to early time behavior (and hence why Gibbs' constant
appears). It is the initial error that slowly decays away as $T$
increases in the integration that plays the important role. 

Given that strategy (2) has error decaying as $1/T$ and (1) as $1/T^{2}$,
the latter is better for long simulations. However, in practice, large
systems and times are inaccessible, i.e., simulations are typically
in the range of 10 to 100 natural time units \citep{Chien13-1,gruss_energy-resolved_2018}.
Thus, the coefficient of the decay matters. Since strategy (1) has
higher error for small $T$, there is a crossing time when strategy
(1) becomes better than (2). This crossing time is much greater than
$100\cdot2\pi/W$ except for $T_{0}=0$, for which it comes at about
$60\cdot2\pi/W$. Thus, averaging within a window (with $T_{0}$ at
an extrema) is generally a better strategy. While these results are
for the specific model under study, many-body systems can display
the same decaying oscillations \citep{Chien13-1,gruss_energy-resolved_2016},
including quantitatively in a large regime of many-body interaction
induced transport \citep{Chien13-1} (which shows the Gibbs phenomenon
and rapid develop of the steady state). Indeed, we closely followed
(2) for many-body transport simulated with matrix product states \citep{gruss_energy-resolved_2016},
albeit empirically determined.

We emphasize that strategy matters, as even if the goal is only moderate
accuracy (e.g., 1 \%), different strategies can mean orders of magnitude
longer simulations requiring an order of magnitude larger system,
as the maximum simulation time\footnote{Equation \eqref{eq:CurrentFinite} has contributions that recur at
a time $\pi N/W$ -- in this case taken as the time, $t\cdot W/N=\pi$,
that a positive contribution turns negative, where $W/N$ is the level
spacing. In natural time units ($2\pi/W$), this time is $N/2$.} is proportional to $N$. If the computational cost scales as $TN^{p}$,
where $p\ge1$, then a 10 times longer simulation will mean at least
a 100 times the computational cost\footnote{For correlation matrices, $p=2$ or $3$ depending on locality of
hopping. For numerical renormalization, formally $p=1$, but longer
simulations generate additional entanglement that gives an effective
$p$ that is larger.}.

The Kubo approach here elucidates the physics behind the development
of the steady state and transient oscillations. These oscillations
are none other than the Gibbs phenomenon due to the filtering of the
current through the electronic bandwidth and bias window. Unlike the
original context of the Gibbs phenomenon \citep{wilbraham_certain_1848,gibbs_fouriers_1898,gibbs_fouriers_1899,bocher_introduction_1906,hewitt_gibbs-wilbraham_1979}
(and in filtering signals), the ringing oscillations are not artifacts,
but physical. For individual particles, the quasi-steady state is
a manifestation of the wave-like nature of particles. However, for
many particles, the current will near its steady state value in time
$\pi/W$. This is why tensor network simulations of the current obtain
reasonable results even for quite small simulations. We expect that
the Kubo approach will assist in understanding other features of simulations,
providing general guidance and informing new strategies for enhancing
efficiency. 

We thank J. Elenewski, M. Ochoa, S. Sahu, C. Rohmann, and P. Haney
for helpful comments.

\end{document}